\documentclass[prl,twocolumn,showpacs,amssymb]{revtex4}

\usepackage{graphicx}
\usepackage{bm}
\def\be{\begin{equation}}
\def\ee{\end{equation}}
\def\bea{\begin{eqnarray}}
\def\eea{\end{eqnarray}}

\begin{document}
\title{Aharonov-Bohm conductance through a single-channel quantum ring:\\
Persistent-current blockade and zero-mode dephasing}
\author{A. P. Dmitriev$^{1,2}$}
\author{I. V. Gornyi$^{1,2}$}
\author{V. Yu. Kachorovskii$^{1,2}$}
\author{D. G. Polyakov$^{2}$}
\affiliation{
$^{1}$A.F.Ioffe Physico-Technical Institute,
194021 St.Petersburg, Russia
\\
$^{2}$Institut f\"ur Nanotechnologie, Karlsruhe Institute of Technology,
76021 Karlsruhe, Germany
}
\date{\today}

\begin{abstract}
We study the effect of electron-electron interaction on transport through a tunnel-coupled single-channel ring. We find that the conductance as a function of magnetic flux shows a series of interaction-induced resonances that survive thermal averaging. The period of the series is given by the interaction strength $\alpha$. The physics behind this behavior is the blocking of the tunneling current by the circular current. The main mechanism of dephasing is due to circular-current fluctuations. The dephasing rate is proportional to the tunneling rate and does not depend on $\alpha$.
\vspace*{-2cm}
\end{abstract}

\maketitle

A major focus of interest in nanophysics \cite{nazarov09} has been quantum interference effects
on one hand and charge-quantization effects on the other, both of which become more prominent
with decreasing dimensionality and size of the device. The prime device for specifically probing
the interference of electrons is a quantum ring connected to the leads. The conductance of the
ring $G(\Phi)$ exhibits the Aharonov-Bohm (AB) effect~\cite{aronov87}, i.e., changes periodically
with the magnetic flux $\Phi$ threading the ring---with a period $\Phi_0=hc/e$---entirely due to
the interference of electron trajectories winding around the hole.

A key concept in the study of coherent electron transport is that of dephasing of electron waves, which at low temperature $T$ is
due to electromagnetic fluctuations produced by electron-electron (e-e) interactions \cite{nazarov09}.
The AB effect is one of the most convenient tools for studying the dephasing processes, since these
directly govern the amplitude of the flux-dependent part of $G(\Phi)$.

The most ideal quantum-ring interferometer would be the one made up of single-channel---ultimately
one-dimensional (1D)---quantum wires. The basic physics of this deceptively simple setup may, however, become conceptually intricate.
Indeed, it is well known that e-e interactions in 1D transform the electron gas into a Luttinger  liquid (LL) \cite{giamarchi04}.
The issue raised is the nature of the interference and dephasing in this
strongly correlated state.
Direct confrontation with experiment appears now to be possible since many-electron nanorings with
a few or single conducting channels have been manufactured \cite{shea00,fuhrer01}.
Transport of interacting electrons through the single-channel ring
is the subject of this paper.

\begin{figure}
\centerline{\includegraphics[width=0.95\columnwidth]{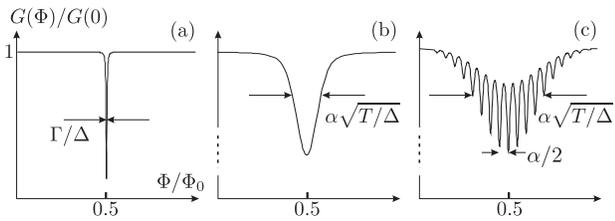}}
\caption{Schematic evolution of $G(\Phi)$ with increasing interaction strength.
(a) $\alpha\ll (\Gamma^2/\Delta T)^{1/2}$: a single deep antiresonance at half-integer flux through the ring; (b) $(\Gamma^2/\Delta T)^{1/2}\ll\alpha\ll \Gamma T/\Delta^2$: suppression of the antiresonance; (c) $\alpha\gg\Gamma T/\Delta^2$: breaking up of the antiresonance into ``persistent-current blockade" oscillations. For fixed $\alpha$, the evolution with increasing $T$ follows $\rm (a)\to (c)\to (b)$.}
\label{f1}
\end{figure}

We study the AB conductance of a LL ring {\it weakly} coupled by tunneling contacts to the leads. Throughout the paper we focus on the high-temperature regime,
\be
T\gg\Delta\gg \Gamma,
\label{0}
\ee
where $\Delta$ is the level spacing inside the ring, $\Gamma$ is the tunneling rate.
Our findings are summarized in Fig.~\ref{f1}. The evolution of $G(\Phi)$ with increasing interaction constant $\alpha$
is governed by two effects specific to the single-channel setup:
(i) the destructive interference at $\Phi=\Phi_0/2$, inherited from the noninteracting problem \cite{buettiker84}, and
(ii) a peculiar type of interaction of electrons with the circular current inside the ring, which dramatically changes the {\it shape} of the interference pattern (Fig.~\ref{f1}).

The physics behind this behavior
can be outlined as follows. The interplay of (i) and (ii)
manifests itself
already in an isolated ring. The interaction with the persistent current $J$ (quantized due to charge quantization)
leads to a shift $\delta\Phi_J\propto \alpha J$  of the effective
flux acting on electrons.
This results
in the interference-induced blocking of the tunneling current through the ring for
specific values of $\Phi$ determined by the quantized values of $J$. We call this phenomenon Persistent-Current Blockade (PCB).

In a tunnel-coupled ring, the circular current $J$ is no longer strictly conserved.
Its dynamics (``zero-mode fluctuations")
is responsible for both the peculiar shape of $G(\Phi)$ and the AB dephasing.
The novel type of interaction-induced oscillations of $G(\Phi)$ that we predict (Fig.~\ref{f1}c)---with a distance between minima controlled by $\alpha$---arises as a series of the PCB antiresonances, each of which corrresponds to one of the quantized values of $J$.
The PCB oscillations---in contrast to the Coulomb-blockade oscillations \cite{nazarov09}---survive thermal averaging at large $T$ but are suppressed by dephasing. As shown below, the dominant mechanism of dephasing in a single-channel ring is provided by thermal fluctuations of the circular current (which translates into fluctuations of $\delta\Phi_J$).
Our main result for the dephasing rate is
\be
\gamma_\varphi=4\Gamma T/\Delta~.
\label{1}
\ee

The dephasing is strongly affected by quantization of charge inside the ring: $\gamma_\varphi$ is
seen to vanish for $\Gamma\to 0$. Another remarkable feature of $\gamma_\varphi$ is that it does
not depend on the interaction strength \cite{remark-1}.
We stress that this zero-mode dephasing is qualitatively different from dephasing in the much better studied electronic
Mach-Zehnder interferometer
\cite{mz}, where for {\it nonchiral} arms the
dephasing rate is given by the single-particle decay rate in a homogeneous LL ($\sim \alpha^2 T$
for spinless electrons \cite{lehur02,gornyi05,seelig01}).

Let us specify the model. Since we are interested in the regime $T\ll\Lambda$, where $\Lambda$ is the ultraviolet cutoff (e.g., the Fermi energy), we linearize the electron dispersion relation around the Fermi level \cite{giamarchi04}. The Hamiltonian reads $H=H_{\rm ring}+H_{\rm tun}+H_{\rm leads}$, where $(\hbar=1)$
\be
H_{\rm ring}=\sum_\mu\int_0^L\!dx\,\left(-i\mu v\psi^\dagger_\mu D_x\psi_\mu+{1\over 2}V_0{\hat n}_\mu {\hat n}_{-\mu}\right)
\label{3}
\ee
describes the isolated LL ring $(D_x=\partial_x-2\pi i\phi/L$, $\phi=\Phi/\Phi_0$).
In this paper we focus on the case of spinless electrons.
The index $\mu=\pm$ denotes electrons moving clockwise (+) and counterclockwise ($-$), $L$
is the circumference of the ring, $V_0$ the zero-momentum Fourier component of the interaction
potential, ${\hat n}_\mu=\,:\!\!\psi^\dagger_\mu\psi_\mu\!\!:$ the density in the channel $\mu$.
We assume that 
the Coulomb interaction is screened by a ground plane
and take the interaction to be point-like. The repulsion between electrons with the same
$\mu$ is then accounted for completely in the renormalization of the velocity
$v$ \cite{gornyi05}.
We characterize the interaction strength by the parameter $\alpha=V_0/2\pi v$.

The tunneling term
$H_{\rm tun}=t_0\big[\psi^\dagger_L\psi(0)+\psi^\dagger_R\psi(L/2)\big]+{\rm h.c.},$
connects  the electron operators $\psi_{R}$ ($\psi_{L}$) in the right (left) lead
at the points of the contacts and $\psi(x)=\psi_+(x)+\psi_-(x)$. The tunneling occurs
at $x=0$ and $L/2$, so that the arms of the interferometer are of the same length.
We consider a symmetric setup with both contacts having the same tunneling rate $\Gamma_0=8\pi|t_0|^2\rho/L$,
where $\rho$ is the (structureless) density of states in the leads at the points of the contacts.
Here we assumed that the leads are noninteracting and ballistic;
the exact form of $H_{\rm leads}$ describing the leads is then of no importance.

In the absence of interaction (see Appendix), the transmission coefficient ${\rm T}(\epsilon,\Phi)$ through the tunnel-coupled ring
shows a resonance \cite{buettiker84} each time the energy $\epsilon$ is close to one of the
eigenenergies $\epsilon_{n\mu}=(n-\mu\phi)\Delta$ of an isolated ring.
At zero $T$ this yields a double-resonance structure in $G(\Phi)$ \cite{buettiker84}.
In the LL ring at $T\ll \Delta$, the AB resonances are
affected by Coulomb blockade and spin-related effects \cite{jagla93,kinaret98,pletyukhov06,eroms08}.

What does not seem to have been generally appreciated in the literature is the behavior of the ``noninteracting"
Landauer conductance
$G_0(\Phi)=(e^2/2\pi)\!\int\!d\epsilon\,(-\partial_\epsilon f)\,
{\rm T}(\epsilon,\Phi) $
in the limit of high temperature $T\gg\Delta$ ($f$ is the thermal distribution function).
Of special interest are the points of degeneracy between levels of different chirality $\mu$
that occur at integer and half-integer values of $\phi$.
At $\phi=1/2$
(which corresponds to the crossing of levels of {\it different} ``parity"),
\be
G_0(\Phi)={e^2\Gamma_0\over 2\Delta}\frac{\cos^2 (\pi\phi)}{\cos^2 (\pi\phi)+(\pi\Gamma_0/2\Delta)^2}
\label{5}
\ee
exactly vanishes. At $\Gamma_0\ll \Delta$, the high-$T$ conductance exhibits a sharp (anti)resonance (Fig.~\ref{f1}a) \cite{remark6}.
By contrast, the interference contribution vanishes at integer $\phi$, where $G_0(\Phi)$ is featureless.

To obtain this behavior in a way that is convenient for introducing
interaction,
let us write ${\rm
T}(\epsilon,\Phi)=|t_{+}(\epsilon,\Phi) +
t_{-}(\epsilon,\Phi)|^2$, where
$t_\pm$ is the transmission amplitude of electrons injected into the $\psi_\pm$ mode.
Multiple returns of electrons to the tunneling contacts described by a 3$\times$3 $S$-matrix are accompanied by changing chirality.
Importantly, at
$\phi=1/2$, for each path contributing to $t_{+}$ there exists a ``mirrored" path (with $\mu\to -\mu$ on each segment) whose contribution to $t_-$ has an opposite sign.
It is this destructive
interference that leads to the vanishing \cite{buettiker84} of ${\rm
T}(\epsilon,\Phi_0/2)$ for arbitrary $\epsilon$.
More specifically, at high $T\gg\Delta$, only the products of amplitudes corresponding to paths of
equal length (but with an arbitrary sequence of chiralities)
are not suppressed by thermal averaging. The conductance can then be written as a sum over the winding numbers $n\geq 0$. A delicate point here is that one cannot neglect backscattering inside the ring at the tunneling contacts even if $\Gamma_0/\Delta$ is small. Doing so would give $G_0(\Phi)\propto \sum_n|A^+_{2n+1}+A^-_{2n+1}|^2$, where $A^{\mu}_k=e^{i \mu k\pi \phi}(1+\pi\Gamma_0/2\Delta)^{1-k}$ is the amplitude that preserves
the chirality of the injected wave (below ${\bar A}^{\mu}_k$ is its complex conjugate). This expression contains sharp resonances {\it both} at $\phi=0$ and at $\phi=1/2$. In fact, however, the effect of backscattering is strongly enhanced by multiple returns and leads to
$G_0(\Phi)\propto\sum_{n\mu}\left[\,|A^\mu_{2n+1}|^2
+\left(A_{2n+1}^\mu{\bar A}^{-\mu}_{2n+1}-A_{2n+2}^\mu{\bar A}^{-\mu}_{2n+2}\right)/2\right]$.
It is seen that the backscattering removes the resonance at $\phi=0$ while not affecting the resonance at $\phi=1/2$.

Our purpose here is to understand how the shape of the AB resonance (\ref{5})
changes when e-e interactions are turned on. Making use of the scale separation (\ref{0}), we first integrate out all energy scales larger than $T$, which takes into account the virtual processes \cite{remark3} that yield the LL renormalization of the model. The main outcome is the renormalization of the tunneling rate: $\Gamma_0\to\Gamma(T)$; in particular,
$\Gamma(T)\sim\Gamma_0(\Lambda/T)^{(1-K)^2/2K}$
for $\alpha\gg\Gamma_0/\Delta$ \cite{unpublished}, where
$K=(1-\alpha)^{1/2}(1+\alpha)^{-1/2}$ is the Luttinger constant.
Note that at $T\gg\Delta$ two contacts are renormalized independently. Another consequence is that the velocity of single-particle excitations \cite{gornyi05} becomes equal to the plasmon velocity $u=v(1-\alpha^2)^{1/2}$ (the renormalized level spacing is now $\Delta=2\pi u/L$).
Next, we employ the quasiclassical approximation---justified for $T\gg\Delta$ and $\alpha\ll 1$---in which the effect of e-e interactions on the single-particle transmission amplitudes is described in terms of scattering on the thermal electromagnetic noise created by the bath of other electrons.

It is instructive to first consider the bath with
the total number $N_\mu$ of electrons in the channel $\mu$ being a quantum number. For a linear dispersion relation,
the peculiarity of the single-channel ring is that at $\Gamma=0$ the density profile $n_{\mu}(x)$ for a given chirality
remains unchanged and rotates
as a whole.
The forward scattering of electrons of chirality $\mu$ is then fully accounted for through the
phase they acquire in the time-dependent potential $U_{\mu}(x,t)= V_0 n_{-\mu}(x+\mu u t)$.
In particular, the quasiclassical amplitude of the
transition from $x=0$ to $x=L/2$ without winding around the hole is given by
$ A^\mu_1=
\exp\Big\{i\pi \mu \phi+i V_0 \int_0^{L/2u}\!\!\! dt\,  n_{-\mu}[x(t)+\mu u t] \Big\}.$
A crucial point is that, even though the time integration is taken over the {\it half}-period, for $x(t)=\mu u t$ the integral is insensitive to a particular profile of $n_{-\mu}$ and only depends on $N_{-\mu}$.
Clearly, this holds true for the amplitude with an arbitrary winding number $n$.
As a result, the interference term in the conductance,
\begin{equation}
A^+_k {\bar A}^-_k= \exp\{2\pi i k[\phi - \alpha(N_+-N_-)/2]\}~,
\label{7}
\end{equation}
is not suppressed by thermal averaging over fluctuations of $n_{\pm}(x,t)$ at fixed $N_\pm$ (it is this averaging that is responsible for the exponential
decay of single-particle excitations in an infinite LL). In other words, plasmons in the isolated ring do not lead
to any dephasing in our symmetric setup.

It follows from Eq.~(\ref{7}) that, apart from the renormalization of $\Gamma$ and $\Delta$,
the only effect of the interaction of electrons tunneling through the ring
with the bath characterized by fixed $N_\mu$
is the effective shift of the flux
\begin{equation}
\delta\Phi_J=-\alpha J \Phi_0/2~,
\label{8}
\end{equation}
where $J=N_+-N_-$ is
the (dimensionless) persistent current circulating inside the ring.
Physically,
the phase shift (\ref{8}) between two interfering waves
stems from the absence of e-e scattering within the same channel $\mu$ (``Hartree-Fock cancellation" \cite{gornyi05}). In effect, for given $J$, electrons of opposite chirality see different electrostatic potentials, which translates into the phase difference in Eq.~(\ref{7}).
Being inserted in Eq.~(\ref{5}), $\delta\Phi_J$ yields a shift of the AB resonance: the PCB occurs at $\phi=1/2-\delta\Phi_J/\Phi_0$; in other words, the persistent current completely blocks the tunneling current through the ring at this value of $\phi$.

For a thermodynamic ensemble of the ``isolated baths", the conductance [Eq.~(\ref{5})] should be averaged  over the Gibbs distribution of the zero-mode energies \cite{zm,remark5},
\begin{equation}
\epsilon_{N_+N_-}=(\Delta/4K)\left[\left(N-N_0\right)^2+K^2\left(J-2\phi\right)^2\right],
\label{8a}
\end{equation}
where
$N_0$ is controlled by the chemical potential and $N=N_++N_-$ is the total number of electrons in the ring. Equation (\ref{8a}) describes, quite generally, electrostatics of a 1D ring. The resulting conductance as a function of $\phi$ shows PCB oscillations with a period $\alpha$ and a Gaussian envelope whose width $w_T=\alpha(T/\Delta)^{1/2}$ is entirely determined by the statistical weights of different values of $J$.

Taking into account the ergodic tunneling dynamics of the electron bath, i.e., the time dependence of the circular current, leads to PCB oscillations in a {\it single} ring \cite{remark4}.
In contrast to the isolated ring, each PCB resonance
acquires a width induced by a finite lifetime of the state with given $J$. Importantly, this time is much shorter than the single-electron tunneling lifetime $\Gamma^{-1}$.
Indeed, the time scale for changing $J$ by unity is given by $\Gamma^{-1}$ divided by the number of levels $T/\Delta$ around the Fermi level that participate in the tunneling dynamics.
We identify the interaction-induced broadening of the PCB resonances with dephasing [Eq.~(\ref{1})].

For a quantitative analysis of $G(\Phi)$, we average the product of the amplitudes in Eq.~(\ref{7}) over realizations of
$J(t)$. This gives the interaction-induced action $S(t_n)$, where  $t_n=2\pi(n+1/2)/\Delta$ for the winding number $n$:
\begin{equation}
e^{-S(t)}=\left\langle\exp\left\{-i\alpha\Delta\int_0^{t}dt^\prime [N_+(t^\prime)-N_-(t^\prime)]\right\} \right\rangle~.
\label{9}
\end{equation}
We now represent $N_\mu=\sum_j n_j^\mu$ as a sum over individual energy levels inside the ring \cite{remark1}.
The time evolution of the occupation numbers $n_j^\mu=0,1$ is telegraph noise with the
rates $\Gamma f_j$ and $\Gamma(1-f_j)$ for scattering ``in" and ``out", respectively, where $f_j$ is the distribution function in the leads at the energy of the $j$th level.
The phase factor induced by the interaction with the $j$th level is written as
(here we suppress the indexes $j$ and $\mu$ for brevity) \cite{qubit}:
$$
\Big\langle e^{i\alpha\Delta \int_0^{t} dt^\prime\, n(t^\prime)}\Big\rangle=(1-f)\left(P_{00}+P_{01}\right)+f\left(P_{10}+P_{11}\right),
$$
where
$P_{kl}(t)$ satisfy the master equation
$\dot{P}_{kl}
=(-1)^l\left\{[\Gamma (1- f)-il\alpha\Delta] P_{k1}-\Gamma fP_{k0}\right\}$
and the initial condition $P_{kl}(0)=\delta_{kl}$ ($k$ and $l$ are the initial and final occupation numbers, respectively).
Solving this equation
we get
$S(t)=-2{\rm Re}\sum_j \ln\left[\left(e^{\lambda^+_j t}\lambda^-_j
- e^{\lambda^-_j t}\lambda^+_j\right)/\left(\lambda^-_j-\lambda^+_j\right)\right],$
where $\lambda^\pm_j=\lambda-i\alpha\Delta f_j\pm(\lambda^2+i\alpha\Delta\Gamma f_j)^{1/2}$ and $\lambda=(i\alpha\Delta-\Gamma)/2$.
The interference term $\delta G(\Phi)=G(\Phi)-G(0)$
is affected by the
action (\ref{9}) (below $\delta_\phi=\phi-1/2$):
\begin{equation}
\frac{\delta G(\Phi)}{G(0)}\simeq -\frac{2\pi\Gamma}{\Delta}\sum_{n=0}^{\infty}
\cos(2\Delta\delta_\phi t_n)\, e^{-\Gamma t_n-S(t_n)}~.
\label{13}
\end{equation}

For
$\alpha\gg\Gamma/\Delta$, the sum in Eq.~(\ref{13}) is cut off by $S(t)$ at $t_n\ll \Gamma^{-1}$, so that we
can expand $S(t)$ in $\Gamma$. The action at $\Gamma=0$ is given by the thermodynamic average $e^{-S_0(t)}=\left<e^{-i\alpha J\Delta t}\right>_{\rm Gibbs}$ over the zero-mode energies (\ref{8a}) and yields PCB resonances with different $J$. For $T\gg\Delta$, $S_0(t)\simeq \alpha^2T\Delta \{t^2\}$
where $\{\ldots\}$ denotes a periodic continuation in $t$ from the interval $-\pi/\alpha\Delta<t<\pi/\alpha\Delta$. The linear-in-$\Gamma$ term,
\begin{equation}
 S_1(t)\simeq\frac{4\Gamma T }{\Delta}\eta (t)\left[t\cos^2\left({\frac{\alpha\Delta t}{2}}\right)
-\frac{\sin (\alpha \Delta t)}{\alpha \Delta }\right]
\label{14}
\end{equation}
with $\eta (t)=\alpha\Delta \{t\}/\sin (\alpha\Delta t)$, is responsible for the dephasing.
For $\alpha\ll (\Delta/T)^{1/2}$, the sum in Eq.~(\ref{13})
can be replaced by an integral. The latter
is dominated
by the vicinity of the points $t=2\pi m/\alpha\Delta$ with integer $m\geq 0$, where $e^{-S_0(t)}$ is sharply peaked. At these points for $m\gg 1$,
$S_1(t)\simeq \gamma_\varphi t$
with the dephasing rate $\gamma_\varphi$ given by Eq.~(\ref{1}). The
interference term
then reads:
\begin{equation}
\frac{\delta G(\Phi)}{G(0)}\simeq {\rm Im}\,
\frac{(\Gamma /2w_T\Delta)\exp(-\delta_\phi^2/w_T^2)}{\sin [\,\pi(\delta_\phi+2i\gamma_\varphi/\Delta)/\alpha\,]}~,
\label{15}
\end{equation}

If $\alpha\Delta \gg \gamma_\varphi$, Eq.~(\ref{15}) yields well-separated Lorentzians \cite{remark7} (Fig.~\ref{f1}c) of width $\gamma_\phi/\Delta$, centered at integer $\delta_\phi/\alpha$. Note that, despite the appearance of the PCB fine structure, the exact period in $\phi$ remains unity, as it should be. In the opposite limit,
$\alpha\Delta \ll \gamma_\varphi$, the broadening of the resonances is larger than the distance between them, so that they merge into a single Gaussian dip of width $w_T$ (Fig.~\ref{f1}b).
Equation (\ref{15}) describes the physically most transparent case of not too large $\alpha\ll (\Delta/T)^{1/2}$, which means that the width $w_T$ of the envelope of the PCB resonances is much smaller than the period of the AB oscillations. At larger $\alpha$, additional features appear; in particular, related to a possible commensurability between $\delta\Phi_J$ and $\Phi_0$---these will be considered elsewhere \cite{unpublished}.

It is worth noting that the tunneling broadens also the plasmon levels inside the ring,
which introduces an additional contribution $\gamma_\varphi^{\text{p}}$ to the dephasing rate. Averaging
the amplitudes $A_k^\mu$ over fluctuations of $n_\mu [x(t)]$ that occur on the time scale of $\Gamma^{-1}$,
we find $\gamma_\varphi^{\text{p}}\sim \alpha^2\Gamma T/\Delta$. It follows that for $\Gamma/\Delta\ll\alpha\ll 1$ the dephasing due to the non-Gaussian zero-mode fluctuations of $J(t)$ is much stronger than that induced by plasmons.

Before concluding, let us briefly discuss an alternative approach \cite{unpublished} to the description of the dephasing in the almost closed geometry, similar to that used to study the full-counting statistics \cite{nazarov09,levitov96}. Within this approach, we represent the interaction-induced action through the time-ordered average
$e^{-S(t)}=\left<\hat{U}_+(t)\hat{U}_-(t)\right>$, where $\hat{U}_\mu (t)=\hat{{\cal T}}e^{-i\mu\alpha\Delta\int_0^t\!dt'\hat{N}_\mu(t')}$ and
$\hat{N}_\mu$ is the operator of the number of electrons inside the ring in the channel $\mu$. The averaging, taken over the equilibrium density matrix for the total Hamiltonian {\it including} the leads, involves a trace operation over the many-particle Fock space. We then use the exact representation of the {\it many}-particle trace in terms of a determinant in the {\it single}-particle space \cite{klich03}: $e^{-S}=\det (1-f+u_+u_-f)$, where $u_\mu$ are now matrices in the basis of scattering states. Importantly, this determinant can be calculated in a parametrically exact way by using the resonance character of scattering [Eq.~(\ref{0})]. The result coincides \cite{unpublished} with Eq.~(\ref{15}), which provides a well-controlled substantiation for the more transparent quasiclassical derivation we chose in this paper.

In conclusion, we have demonstrated that e-e interactions lead to profound and unusual effects in transport through a single-channel quantum-ring interferometer tunnel-coupled to the leads, originating from the phenomenon of Persistent-Current Blockade. We have shown that the AB conductance $G(\Phi)$ exhibits a series of sharp resonances broadened by dephasing, the distance between which is controlled by the interaction strength. We have calculated the main contribution to the dephasing rate, which is due to tunneling-induced fluctuations of the circular current. The physics described in the paper remains intact for spinful electrons and ballistic systems with a small number of conducting channels. Our predictions can thus be verified by measuring the conductance of a semiconductor nanoring or a single coil of carbon nanotube.

We thank D.~Aristov, D.~Bagrets, H.~Bouchiat,
Y.~Imry, D.~Khmelnitskii, V.~Khrapai, A.~Mirlin, M.~Pletyukhov, and Y.~Stein
for valuable discussions.
The work was supported by the DFG/CFN, the EUROHORCS/ESF EURYI Award, GIF Grant No.\ 965, the RFBR, Programs of the RAS, the Dynasty Foundation, and Rosnauka
Grant 02.740.11.5072.

\begin{widetext}
\section {Appendix: Noninteracting electrons}
\label{aA}
\renewcommand{\theequation}{A.\arabic{equation}}
\setcounter{equation}{0}

In this appendix, we briefly discuss the derivation of Eq.~(\ref{5}) and the modification of the latter to the case of an asymmetric setup.
We start from the expression for the transmission coefficient ${\rm T}(k,\phi,a)$ (the transmission amplitude squared) of an electron with energy $\epsilon=v_Fk$ for the ring interferometer having two arms with the lengths $(L-a)/2$ and $(L+a)/2$. For a symmetric setup, ${\rm T}(k,\phi,a=0)$ was obtained earlier in Ref.~\cite{buettiker84} (see also Refs.~\cite{jagla93,pletyukhov06}). A straightforward generalization to the case of $a\neq 0$ yields:
\begin{equation}
{\rm T}(k,\phi,a)
=\frac{8 \gamma^2\{2\cos^2(\pi\phi)\sin^2(kL/2)-[\,\cos (2\pi\phi)-\cos (kL)\,]\sin^2(ka)\}}
{\{\cos (2\pi\phi)-\cos (kL)+2\gamma^2 [\,\sin^2 (kL/2) - \sin^2(ka)\,]\}^2 +4 \gamma^2 \sin^2 (kL)}~,
\label{A1}
\end{equation}
where $\gamma=\pi\Gamma_0/2\Delta$.
For $\gamma \ll 1$, Eq.~(\ref{A1}) predicts sharp resonances at $kL=2\pi(n\pm \phi )$, where $n$ is an integer.
The conductance is obtained by thermal averaging of Eq.~(\ref{A1}) over $k$:
\begin{equation}
G=\frac{e^2}{2\pi}\left \langle {\rm T}(k,\phi,a) \right \rangle_T~.
\label{A2}
\end{equation}

We perform the averaging in Eq.~(\ref{A2}) in two steps.
We first expand Eq.~(\ref{A1}) in a series in $\exp(ikL)$ (the coefficients of this expansion still
depend on $k$ through the parameter $ka$) and notice that all harmonics except the zeroth one can be
neglected for $T\gg\Delta$ and $a\ll L$.
After some cumbersome algebra we arrive at the following expression for the zero harmonic:
\begin{equation}
G=\frac{e^2 \gamma}{\pi} \left \langle
\frac{ \sin^2 (\pi \phi)  \sin^2 ({ka}/{2})  }{ \sin^2 (\pi \phi) + \gamma^2 \sin^2 ({ka}/{2})}
+\frac{ \cos^2 (\pi \phi)  \cos^2 ({ka}/{2})  }{ \cos^2 (\pi \phi) + \gamma^2 \cos^2 ({ka}/{2})} \right \rangle_T~.
\label{A3}
\end{equation}
At the second step we do the remaining thermal averaging in Eq.~(\ref{A3}), which yields the dependence of $G$ on the parameters $k_Fa$ and $Ta/v_F$.

In the symmetric setup, the first term in Eq.~(\ref{A3}) vanishes, while the second one gives Eq.~(\ref{5}). For $\gamma \ll 1 $, the conductance as a function of $\phi$ in the symmetric ring has a periodic sequence of
sharp antiresonances at half-integer values of $\phi$. The depth of the antiresonance is unity (in units of $e^2\gamma/\pi$)
and the width is $\gamma/\pi$. The shape of the antiresonance is given by a simple Lorentzian:
\begin{equation}
G\simeq {e^2\gamma\over\pi}\frac{ \delta_\phi^2}{\delta_\phi^2 +\gamma^2/\pi^2}~,
\label{A4}
\end{equation}
where $\delta_\phi$ is the distance from the center of the antiresonace.

Let us now turn to the asymmetric setup ($a\neq 0$). We first consider the case $Ta/v_F \gg 1$.
In this limit, the thermal averaging in Eq.~(\ref{A3}) reduces to the integration over the phase $ka$ in the interval $-\pi<ka<\pi$. The result reads:
\begin{equation}
G= \frac{e^2 {\gamma}}{\pi} \left[\,F\left(\cos{\pi \phi}\right)+F\left(\sin{\pi\phi}\right)\,\right]~,
\label{A5}
\end{equation}
where the function $F(z)$ is given by
\begin{equation}
\label{A6}
F(z)=\frac{ z^2 }{ \sqrt{z^2+\gamma^2} \left(|z|+\sqrt{z^2+\gamma^2} \right) }~.
\end{equation}
As follows from Eqs.~(\ref{A5}) and (\ref{A6}), for $\gamma \ll 1 $ the conductance has sharp antiresonances at {\it both} half-integer and integer values of $\phi$. The depth of all of the antiresonances is 1/2 (in units of  $e^2\gamma/\pi$) and the width is $\gamma/\pi$. In contrast to Eq.~(\ref{A4}), the conductance is equal to $1/2\times (e^2\gamma/\pi)$ (not zero!) at the antiresonance points and the shape of the dips is more complicated than the Lorentzian:
\begin{equation}
G\simeq  {e^2\gamma\over\pi}\left[\,\frac{1}{2}+\frac{ \delta_\phi^2}{\sqrt{\delta_\phi^2 +\gamma^2/\pi^2}\left(|\delta_\phi|+\sqrt{\delta_\phi^2+\gamma^2/\pi^2}\right)}\,\right]~.
\label{A7}
\end{equation}
Thus, in the strongly asymmetric case with $Ta/v_F\gg 1$, the amplitude of the dips at half-integer fluxes decreases as compared to the symmetric case from 1 to 1/2, while the additional dips with the amplitude 1/2 appear at integer fluxes.

The dependence of the conductance on $\phi$ becomes even more intricate for $Ta/v_F \ll 1$. In this case, the conductance is given by Eq.~(\ref{A3}) with $ k$ replaced by $k_F$:
\begin{equation}
G=\frac{e^2\gamma}{\pi}\left[\,\frac{\sin^2(\pi\phi)\sin^2({k_Fa}/{2})}{\sin^2 (\pi\phi)+\gamma^2 \sin^2({k_Fa}/{2})}
+\frac{\cos^2(\pi\phi)\cos^2 ({k_Fa}/{2})}{\cos^2(\pi\phi)+\gamma^2 \cos^2 ({k_Fa}/{2})}\,\right]~.
\label{A8}
\end{equation}
For $\gamma\ll 1$, Eq.~(\ref{A8}) shows two series of narrow antiresonances: at integer fluxes, where they are described by
\begin{equation}
G\simeq {e^2\gamma\over\pi}\left[\,\cos^2(k_Fa/2)+\frac{\delta_\phi^2\sin^2(k_Fa/2)}{\delta_\phi^2 +{(\gamma/\pi)^2\sin^2(k_Fa/2)}}\,\right]~,
\label{A9}
\end{equation}
and at half-integer fluxes, where the antiresonance is given by
\begin{equation}
G\simeq {e^2\gamma\over\pi}\left[\,\sin^2(k_Fa/2)+\frac{\delta_\phi^2\cos^2(k_Fa/2)}{\delta_\phi^2+{(\gamma/\pi)^2 \cos^2 (k_Fa/2)}}\,\right]
\label{A10}
\end{equation}
(in both equations $\delta_\phi$ means the deviation from the antiresonance). It is seen that both the amplitudes and the widths of the dips oscillate with changing parameter $k_Fa$.

It is clear that the dependence of the conductance on $a$ is related to the oscillatory behavior of the product of two amplitudes which propagate in opposite directions around the hole and acquire the phase difference $ka$. What is less obvious is the dependence of the interference terms on the parameter $Ta/v_F$. Naively, one might think that in the limit of $Ta/v_F\gg 1$ each and every interference term is suppressed by thermal averaging and the conductance is entirely determined by the classical contribution in which the paths for two amplitudes are the same. However, as seen from Eqs.~(\ref{A5})--(\ref{A7}),  $G$ remains a strong function of $\phi$ even for $Ta/v_F\gg 1$, so that some interference contributions do survive in this limit.

To understand this behavior, we observe that there is a special class of interference terms $A_{2n}^\mu\bar{A}_{2n}^{-\mu}$ in which the amplitudes encircle the hole an {\it integer} number of times. The simplest contribution to the conductance of this type is related to the amplitudes that first propagate between the left and right contacts classically by encircling the hole $m$ times, then split at the right contact and start moving in opposite directions, encompass the flux $n$ times, and finally interefere at the right contact. The contribution to the conductance from these processes, $\Delta G$, reads
\begin{equation}
\Delta G\propto {\rm Re}\sum_{nm\mu}|A_{2m+1}^\mu|^2 t_b A_{2n+2}^\mu\bar{A}_{2n+2}^{-\mu}~.
\label{A11}
\end{equation}
One can see that for given $n$ and $m$ the lengths of two interfering paths in this type of scattering processes are equal to each other---hence the interference term survives thermal averaging---and do not depend on $a$. Importantly, the small factor $t_b=-\gamma/2$ in Eq.~(\ref{A11}), which is the amplitude of backscattering inside the ring from the tunneling contact is compensated by the large sum over the {\it classical} returns to the contact: $\sum_{m}|A_{2m+1}^\mu|^2\sim\gamma^{-1}$, and so does not lead to any additional smallness of $\Delta G$ [see also the discussion below Eq.~(\ref{5})].

In the presence of interaction, the contribution $\Delta G$ is not affected by the plasmon dephasing for the same reason as for $a=0$ [cf.\ Eqs.~(\ref{7}) and (\ref{A11})]. Therefore, the physics of the PCB and zero-mode dephasing remains intact in the asymmetric ring.


\end{widetext}


\begin{thebibliography}{99}
\bibitem{remark6} The resonance in the energy-averaged conductance at $\phi=1/2$ was obtained in Ref.~\cite{jagla93}, where the zero-$T$ case was considered but, to present the results, the authors have averaged them over $\epsilon$ in the interval of width $\Delta$.
\bibitem{remark3} We use the scheme of separating the virtual and real processes in a LL that was devised in Ref.~\cite{gornyi05}.
\bibitem{remark5} As pointed out in Ref.~\cite{pletyukhov06}, $G(\Phi)$ cannot be decoupled into a product of two single-particle zero-mode averages.
\bibitem{remark4} The fluctuating circular current induces fluctuations of $\delta\Phi_J$ and thus of the tunneling current: transport is totally blocked within an interval of time when $J(t)$ is such that $\delta\Phi_J(t)$ compensates the deviation of $\phi$ from 1/2.
\bibitem{remark1} We neglect the inelastic scattering of bath electrons on each other
because it does not affect
$N_\mu(t)$.
\bibitem{remark7} Equation~(\ref{15}) can be cast in the form of a Gibbs sum over zero-mode states: $\delta G(\Phi)/G(0)\simeq \Gamma\gamma_\varphi\left<[(2\delta_\phi-\alpha J)^2\Delta^2+\gamma_\varphi^2]^{-1}\right>_{\rm Gibbs}$.
\end{thebibliography}
\end{document}